\documentclass[floatfix,aps,superscriptaddress,prd,amsmath,nofootinbib,preprintnumbers,onecolumn]{revtex4}

\usepackage{graphicx}
\usepackage{bm}
\usepackage{float}
\usepackage[
colorlinks=true,        % color link
citecolor=blue,         % cite color
linkcolor=blue,         % link color
urlcolor=blue           % url color
]{hyperref}             % create hyperlinks

\newcommand{\nc}{\newcommand*}
\nc{\Om}{\Omega}
\nc{\ogw}{\Omega_{\mathrm{GW}}}
\nc{\rd}{\mathrm{d}}
\nc{\eg}{\textit{e.g.~}}
\nc{\red}[1]{\textcolor{red}{#1}}
\nc{\lvc}{LIGO/Virgo} % LIGO-VIRGO collaboration

\def\({\left(}
\def\){\right)}
\def\[{\left[}
\def\]{\right]}

\def\e{\begin{equation}}
\def\q{\end{equation}}
\def\m{\begin{eqnarray}}
\def\n{\end{eqnarray}}

\begin{document}

\title{Measuring the modified gravitational waves propagation beyond general relativity from CMB observations}

%%%%%%%%%%%%%%%%%%%%%%%%%%%%%%%%%%%% author %%%%%%%%%%%%%%%%%%%%%%%%%%%%%%%%%%%%
\author{Jun Li}
\email{lijun@qust.edu.cn}
\affiliation{School of Mathematics and Physics,
    Qingdao University of Science and Technology,
    Qingdao 266061, China}
\affiliation{CAS Key Laboratory of Theoretical Physics,
    Institute of Theoretical Physics, Chinese Academy of Sciences,
    Beijing 100190, China}

\date{\today}

\begin{abstract}
In modified gravity theories, the gravitational waves propagation are presented in nonstandard ways. We consider a friction term different from GR and constrain the modified gravitational waves propagation from observations. The modified gravitational waves produce anisotropies and polarization which generate measurable tensor power spectra. We explore the impact of the friction term on the power spectrum of B-modes and the impact on the constraints on the other parameters (e.g., $r$ or $A_t$) when $\nu_0$ is allowed to vary in the Monte Carlo analyses from Planck+BK18 datasets. If we assume the result of the scalar perturbations is unchanged, the inflation consistency relation alters with the friction term. In the $\Lambda$CDM+$r$+$\nu_0$ model, the tensor-to-scalar ratio and the amplitude of tensor spectrum are influenced obviously. 
\end{abstract}


\maketitle

%%%%%%%%%%%%%%%%%%%%%%%%%%%%%%%%%%%%%%%%
%%%%%%%%%%%%%%%%%%%%%%%%%%%%%%%%%%%%%%%
\section{introduction}
In recent decades, general relativity (GR) has become an integral and indispensable part of modern physics. While GR describes the dynamics of space-time and governs the behaviors of our universe extremely successful, the rapid development of observations inspire us to explore new physic beyond GR. Modified gravity theories provide plausible possibilities which are worth pursuing and could give a better understanding of observation. In modified gravity theories, the evolution equations of gravitational waves are presented in nonstandard ways \cite{Brax:2017pzt,Amendola:2014wma,Xu:2014uba,Lin:2016gve,Raveri:2014eea,Pettorino:2014bka,Ezquiaga:2021ler,Cai:2020ovp,Belgacem:2019zzu,Boubekeur:2014uaa,Dubovsky:2009xk,Bian:2021ini,Li:2017cds}. The friction term is a fundamental issue on the propagation of gravitational waves. When the gravitational waves propagate with a friction term different from GR, this scenario would arise a variety of modified gravity theories \cite{Amendola:2014wma,Xu:2014uba,Lin:2016gve,Pettorino:2014bka,Ezquiaga:2021ler,Belgacem:2019zzu,Boubekeur:2014uaa,Bian:2021ini}. Probing the friction term is an important way to explore modified gravity and underlying new physic. Hence, we need to investigate the possible deviations from GR and search for corresponding observable effects at cosmological scales.

The friction term on gravitational waves propagation is time-dependent variation and represents damping effect. Modified gravity models provide the motivation for the deviations of friction term, such as the Horndeski theories \cite{DeFelice:2011bh, Kobayashi:2019hrl, Bellini:2014fua, Saltas:2014dha}, the nonlocal infrared modification of gravity \cite{Belgacem:2018lbp, Belgacem:2017ihm} or the quantum chromodynamic (QCD) phase transition \cite{Byrnes:2018clq, Hajkarim:2019nbx}. Recently, there are many works to discuss and constrain the friction terms of GW from the LIGO-Virgo Collaboration \cite{Zhao:2019xmm, Mancarella:2021ecn, Ezquiaga:2021ler}. Furthermore, some papers assume a model-independent friction term and search for corresponding signatures. Instead of testing individual modified gravity models, they parametrize and investigate the departures from GR. Parametrization has been used widely in physics, such as the $\Lambda$CDM model which can point the direction of modified gravity by the deviations of parameters. The tensor-mode parametrization for modified gravity has been proposed in \cite{Lin:2016gve} which discuss a general form of the modified tensor-mode propagation including several physical effects. The inflation consistency relation is modified with friction term, but they do not consider this when updating the constraints on friction.

In this paper, we consider the same friction term as \cite{Lin:2016gve} and constrain the modified gravitational waves propagation from observations. When we consider the modified friction term, the behavior of gravitational waves changes. The modified gravitational waves produce anisotropies and polarization which generate measurable tensor power spectra. Here we explore the impact of the friction term on the power spectrum of B-modes and the impact of the constraints on the other parameters (e.g., $r$ or $A_t$) when $\nu_0$ is allowed to vary in the Monte Carlo analyses from Planck observations \cite{Planck:2018vyg} the BICEP/Keck Observations through the 2018 Observing Season (BK18) \cite{BICEP:2021xfz}. Owing to the modified inflation consistency relation, the tensor-to-scalar ratio and the amplitude of tensor spectrum are influenced obviously.

The modified gravity may have an impact on both tensor and scalar perturbations. If~all of the modified terms are considered together, it is hard to figure out the effects of the friction term. Here, we parametrize the friction term only and investigate corresponding observable~effects.

\section{the modified gravitational waves propagation}
In the conformal Newtonian gauge, the metric about the Friedmann-Robert-Walker background is taken as
\e
\mathrm{d}s^2=a^2\left\{-(1+2\Phi)\mathrm{d}\eta^2+\left[(1-2\Phi)\delta_{ij}+\frac{h_{ij}}{2}\right]\mathrm{d}x^i\mathrm{d}x^j \right\},      \label{metric}
\q
where $a(\eta)$ is the scale factor, $\eta$ is the conformal time, $\Phi$ is the scalar perturbation and $h_{ij}$ is the gravitational waves perturbation.
In GR, the gravitational waves satisfy following wave equation
\e
h_{k}^{\prime\prime}+2\frac{a^\prime}{a}h_{k}^\prime+k^2h_{k}=0, \label{gw}
\q
where the prime denotes derivative with respect to conformal time and the source term is ignored. Tensor perturbations produce anisotropies and polarization which could generate measurable tensor angular power spectra. The temperature and polarization perturbations satisfy Boltzmann equations \cite{Zaldarriaga:1996xe}\m
&\tilde{\Delta}_T^{\prime(T)}+ik\mu\tilde{\Delta}_T^{(T)}=-h^{\prime}-\tau^{\prime}[\tilde{\Delta} _T^{(T)}-\Psi], \\
&\tilde{\Delta}_P^{\prime(T)}+ik\mu\tilde{\Delta}_P^{(T)}=-\tau^{\prime}[\tilde{\Delta} _P^{(T)}+\Psi],
\n
where
\m
\Psi=\frac{\tilde{\Delta}_{T0}^{(T)}}{10}+\frac{\tilde{\Delta}_{T2}^{(T)}}{7}+\frac{3\tilde{\Delta}_{T4}^{(T)}}{70}-\frac{3\tilde{\Delta}_{P0}^{(T)}}{5}+\frac{6\tilde{\Delta}_{P2}^{(T)}}{7}-\frac{3\tilde{\Delta}_{P4}^{(T)}}{70},
\n
the variables $\tilde{\Delta}_{T}^{(T)}$ and $\tilde{\Delta}_{P}^{(T)}$ describe the temperature and polarization perturbations generated by gravitational waves,
the superscript $(T)$ denotes contributions from tensor perturbations, $\mu=\hat{n}\cdot\hat{k}$ is the angle between photon direction and wave vector, $\tau^{\prime}$ is the differential optical depth for Thomson scattering. The~multipole moments of temperature and polarization are defined as $\Delta(k,\mu)=\sum_l(2l+1)(-i)^l\Delta_l(k)P_l(\mu)$, where $P_l(\mu)$ is the Legendre polynomial of order $l$.
The polarization perturbations can be decomposed into E-mode and B-mode. The~B-mode components mainly come from tensor perturbations on the small multipoles and contain information about gravitational waves. The~polarization power spectra from tensor perturbations are given by~\cite{Zaldarriaga:1996xe}

\m
C_{X\ell}^{(T)}=(4\pi)^2\int k^2\mathrm{d}kP_h(k)\left|\Delta_{X\ell}^{(T)}(k,\eta=\eta_0)\right|^2\label{power spectrum1}, 
\n
where $P_h(k)$ is the primordial power spectrum of gravitational waves, $X$ stands for $E$ or $B$. The two-point correlations of polarization patterns at different points in the sky are presented by Eq.~(\ref{power spectrum1}). The tensor perturbations could generate measurable BB tensor angular power spectrum.

The power spectrum from tensor perturbations is parameterized as
\m
P_h(k)&=&A_t\(\frac{k}{k_*}\)^{n_t},\label{tensor}
\n
where $A_t$ is the tensor amplitude at the pivot scale $k_*=0.05$ Mpc$^{-1}$, $n_t$ is the tensor spectral index.
In literature, the tensor-to-scalar ratio $r$ is used to quantify the tensor amplitude compared to the scalar amplitude $A_s$ at the pivot scale, namely
\e
r\equiv\frac{A_t}{A_s}.
\q
For the canonical single-field slow-roll inflation model, $n_t$ is related to $r$ by $n_t=-r/8$ which is called inflation consistency relation in general relativity \cite{Liddle:1992wi,Copeland:1993ie}.

In this paper, we suggest the following form of the modified propagation equation for tensor perturbations 
\e
h_{k}^{\prime\prime}+(2+3{\nu_0})\frac{a^\prime}{a}h_{k}^\prime+k^2h_{k}=0, \label{mgw}
\q
where ${\nu_0}$ is a constant parameter. If the background is exactly exponentially expanding with respect to the cosmic time as $a\propto e^{Ht}$, the tensor mode is given by \cite{Lin:2016gve}
\e
\left|h_k^0\right|^2=\frac{G(2H)^{2+3{\nu_0}}\left[\Gamma(\frac{3}{2}+\frac{3}{2}{\nu_0})\right]^2}{\pi^3\cdot k^{3+3{\nu_0}}},
\q
where $h_k^0$ is the leading-order solution, $H$ is the constant expansion rate during inflation and $G$ is the Newtonian constant. The case ${\nu_0}=0$ corresponds to the propagation in GR \cite{Riotto:2002yw}. The power spectrum of gravitational waves is defined as 
\e
P_h(k)=\frac{k^3}{2\pi^2}\left|h_k^0\right|^2.
\q
Comparing with Eq.~(\ref{tensor}), we can identify the tensor spectral index as 
\e
n_t=-3{\nu_0}.
\q
For the slow-roll inflation, $H$ is not a constant and measured by the slow-roll parameter $\epsilon=-\dot H/H^2$. The tensor spectrum index becomes
\e
n_t=-3{\nu_0}-2\epsilon.
\q
If we assume the result of the scalar perturbations is unchanged, the tensor-to-scalar ratio $r$ is still related to the slow-roll parameter as $r=16\epsilon$. The inflation consistency relation in the modified gravity becomes 
\e
n_t=-3{\nu_0}-r/8.\label{mcr}
\q

\begin{figure}[thb!]
\centering
\includegraphics[width=10cm]{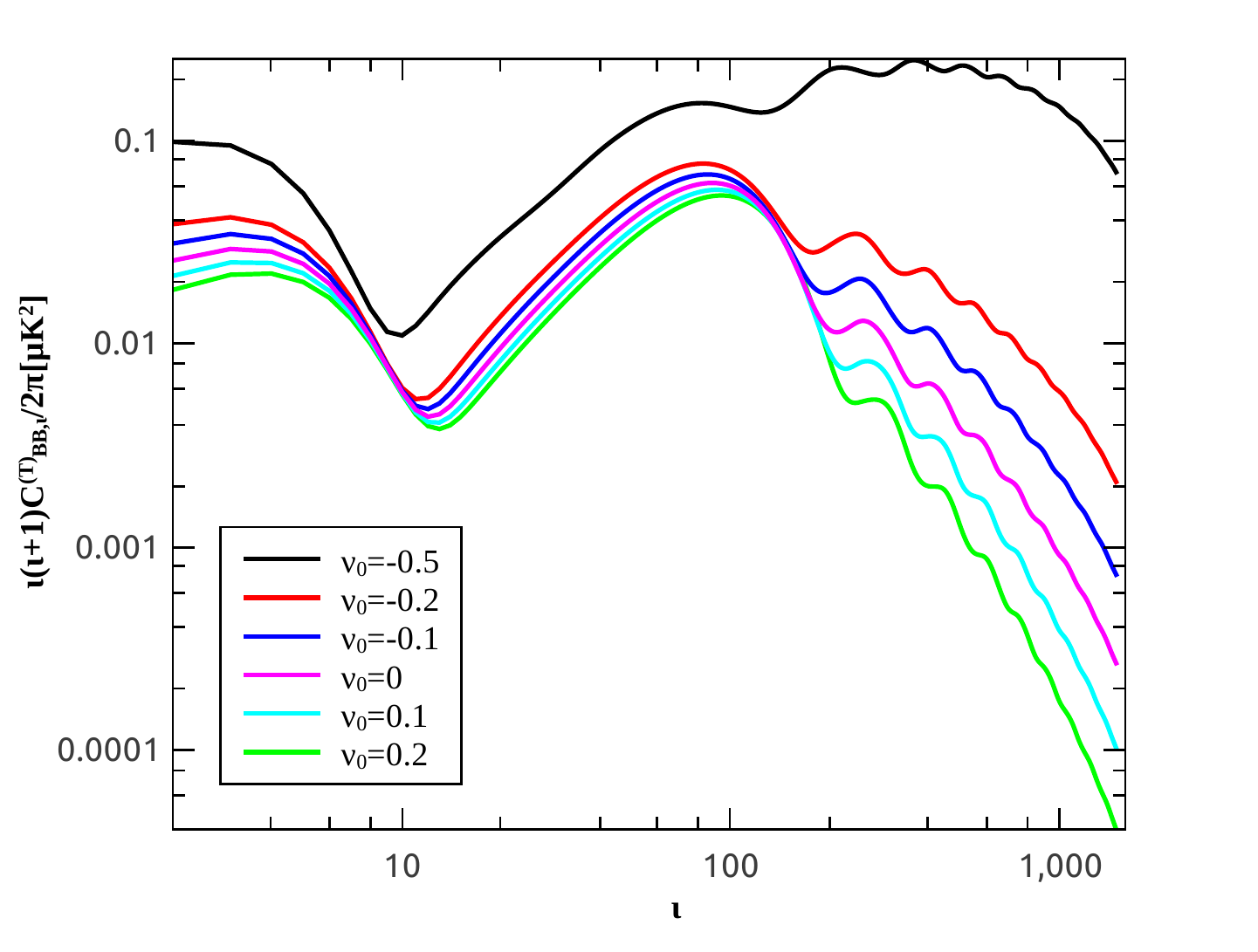}
\caption{The plot of BB angular power spectrum from tensor perturbations for $\nu_0=-0.5$, $\nu_0=-0.2$, $\nu_0=-0.1$, $\nu_0=0$, $\nu_0=0.1$ and $\nu_0=0.2$, respectively.}
\label{fig1}
\end{figure}

In order to obtain the tensor angular power spectra for the modified gravitational waves propagation, we modify CAMB \cite{Hojjati:2011ix} by taking into account the Eq.~(\ref{mgw}). Our numerical results are presented in Fig.~\ref{fig1}. We show that the modified gravitational waves propagation has impacts on BB tensor angular power spectrum. The negative $\nu_0$ enhances the BB tensor angular power spectrum, while the positive $\nu_0$ reduces the BB tensor angular power spectrum.

\section{the constraints on friction term from BK18 data}
\begin{figure}[thb!]
\centering
\includegraphics[width=14cm]{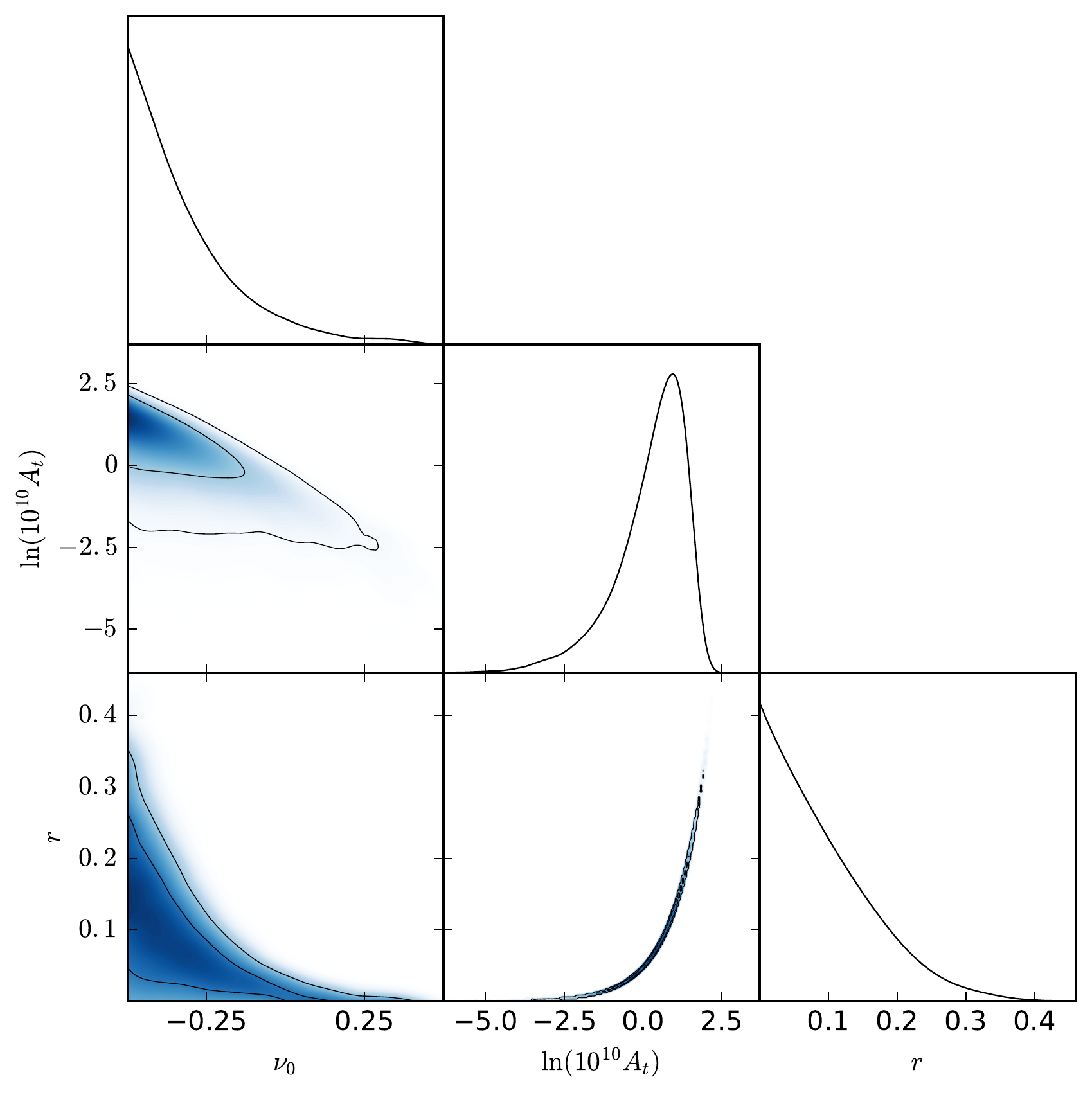}
\caption{The contour plots and the likelihood distributions for parameters $r$, $\nu_0$ and $\ln\(10^{10}A_t\)$ in the $\Lambda$CDM+$r$+$\nu_0$ model at the $68\%$ and $95\%$ CL from Planck+BK18 datasets.}
\label{fig2}
\end{figure}

In the standard $\Lambda$CDM model, the~six parameters are the baryon density parameter $\Omega_b h^2$, the~cold dark matter density $\Omega_c h^2$, the~angular size of the horizon at the last scattering surface 
$\theta_\text{MC}$, the~optical depth $\tau$, the~scalar amplitude $A_s$ and the scalar spectral index $n_s$. We extend this model by adding the tensor-to-scalar ratio $r$ and the friction factor $\nu_0$, and~consider these eight parameters as fully free parameters, i.e.,~$r\in[0, 2]$, $\nu_0\in[-0.5, 0.5]$. We use the publicly available codes Cosmomc~\cite{Lewis:2002ah} to constrain parameters, which adds the modified gravitational waves propagation in Equation~(\ref{mgw}) and the modified inflation consistency relation in Equation~(\ref{mcr}). The~numerical results are presented in Figure~\ref{fig2}.

In the $\Lambda$CDM+$r$+$\nu_0$ model, the constraints on the tensor-to-scalar ratio $r$ and the friction factor $\nu_0$ are
\m
r &<& 0.243 \quad(95\% \ \mathrm{C.L.}),\\
\nu_0&<& 0.042  \quad(95\% \ \mathrm{C.L.}),
\n
from Planck+BK18 datasets. These results show that the friction factor $\nu_0$ refers to negative region and negative $\nu_0$ enhances the upper limits on tensor-to-scalar ratio. We cut out the small $\nu_0$ parameter space, because a smaller $\nu_0$ leads to a larger tensor-to-scalar ratio.

\begin{figure}[thb!]
\centering
\includegraphics[width=11cm]{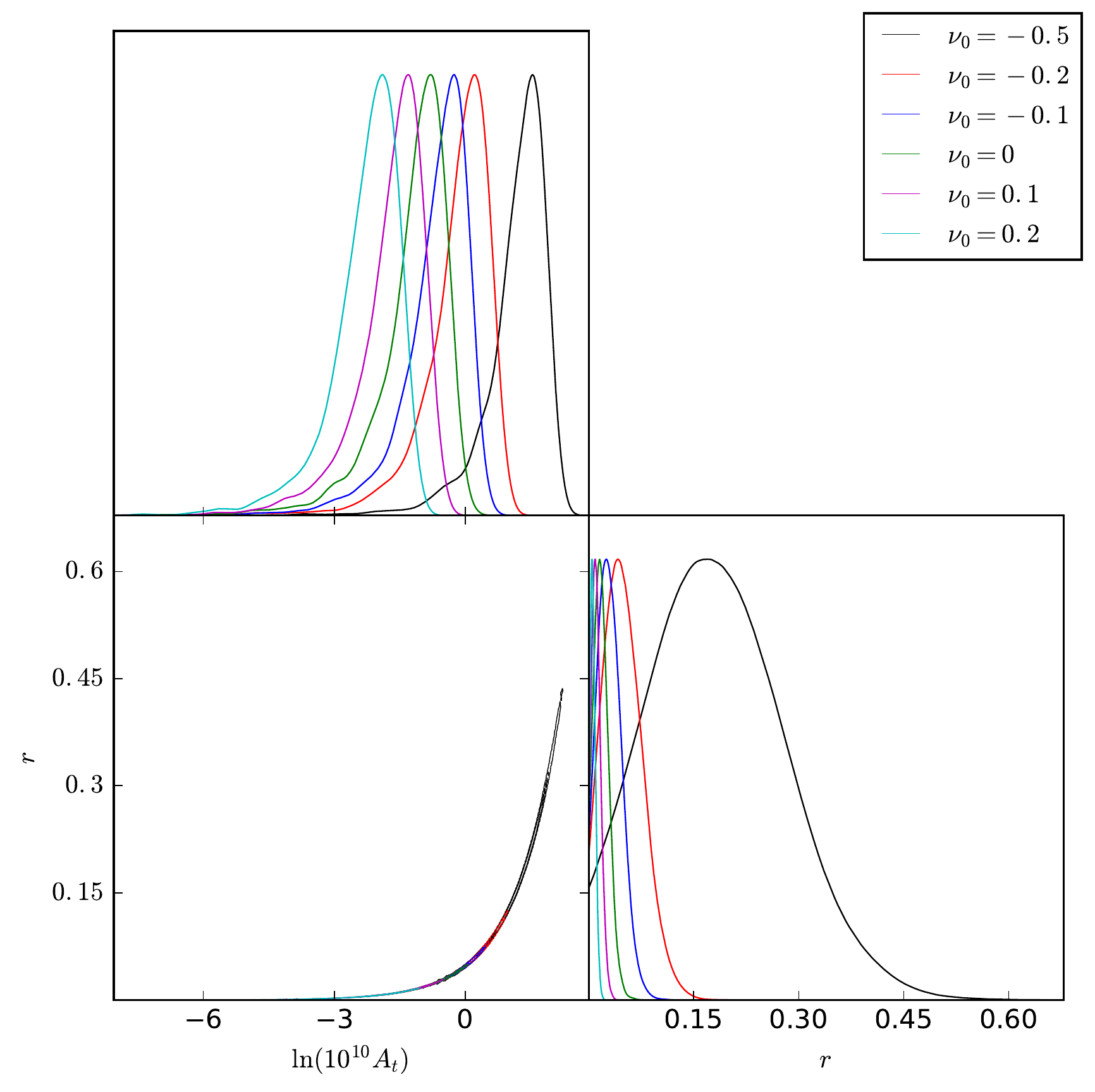}
\caption{The contour plot and the likelihood distributions for parameters $r$ and $\ln\(10^{10}A_t\)$ in the $\Lambda$CDM+$r$+$\nu_0$ model at the $68\%$ and $95\%$ CL from BK18 with the modified gravitational waves propagation for $\nu_0=-0.5$, $\nu_0=-0.2$, $\nu_0=-0.1$, $\nu_0=0$, $\nu_0=0.1$ and $\nu_0=0.2$, respectively.}
\label{fig3}
\end{figure}

Then, we fix the friction factor $\nu_0$ and consider several cases to test the variation on tensor parameters. As~we assume the result of the scalar perturbations is unchanged, we fix the standard $\Lambda$CDM parameters based on Planck observations: $\Omega_b h^2=0.02242$, $\Omega_c h^2=0.11933$, $100\theta_\text{MC}=1.04101$, $\tau=0.0561$, $\ln\(10^{10}A_s\)=3.047$ and $n_s=0.9665$. The~numerical results are presented in Figure~\ref{fig3}. In~the $\Lambda$CDM+$r$+$\nu_0$ model, the~constraints on the tensor-to-scalar ratio $r$ and the tensor amplitude are
\m
r &<& 0.349 \quad(95\% \ \mathrm{C.L.}),\\
\ln\(10^{10}A_t\)&=&1.17^{+0.81}_{-0.34}  \quad(68\% \ \mathrm{C.L.}),
\n
from BK18 for $\nu_0=-0.5$. The constraints on the tensor-to-scalar ratio $r$ and the tensor amplitude are
\m
r &<& 0.100 \quad(95\% \ \mathrm{C.L.}),\\
\ln\(10^{10}A_t\)&=&-0.18^{+0.90}_{-0.37}  \quad(68\% \ \mathrm{C.L.}),
\n
from BK18 for $\nu_0=-0.2$. The constraints on the tensor-to-scalar ratio $r$ and the tensor amplitude are
\m
r &<& 0.060 \quad(95\% \ \mathrm{C.L.}),\\
\ln\(10^{10}A_t\)&=&-0.68^{+0.89}_{-0.38}  \quad(68\% \ \mathrm{C.L.}),
\n
from BK18 for $\nu_0=-0.1$. The constraints on the tensor-to-scalar ratio $r$ and the tensor amplitude are
\m
r &<& 0.037 \quad(95\% \ \mathrm{C.L.}),\\
\ln\(10^{10}A_t\)&=&-1.20^{+0.93}_{-0.38}  \quad(68\% \ \mathrm{C.L.}),
\n
from BK18 for $\nu_0=0$. The constraints on the tensor-to-scalar ratio $r$ and the tensor amplitude are
\m
r &<& 0.023 \quad(95\% \ \mathrm{C.L.}),\\
\ln\(10^{10}A_t\)&=&-1.71^{+0.93}_{-0.40}  \quad(68\% \ \mathrm{C.L.}),
\n
from BK18 for $\nu_0=0.1$. The constraints on the tensor-to-scalar ratio $r$ and the tensor amplitude are
\m
r &<& 0.013 \quad(95\% \ \mathrm{C.L.}),\\
\ln\(10^{10}A_t\)&=&-2.30^{+0.94}_{-0.43}  \quad(68\% \ \mathrm{C.L.}),
\n
from BK18 for $\nu_0=0.2$. We show that the modified gravitational waves propagation and the inflation consistency relation in the modified gravity have impacts on the tensor-to-scalar ratio $r$ and the tensor amplitude. The negative $\nu_0$ enhances the upper limits on the tensor-to-scalar ratio $r$, while the positive $\nu_0$ reduces the upper limits. The related behaviors of tensor-to-scalar ratio $r$ and the friction factor $\nu_0$ are in agreement with the variations in Fig.~\ref{fig2}.

\section{summary}
In this paper, we consider a friction term different from GR and constrain the modified gravitational-wave propagation from observations. We consider the impact of the friction term on the power spectrum of B-modes and the impact of the constraints on the other parameters (e.g., $r$ or $A_t$) when $\nu_0$ is allowed to vary in the Monte Carlo analyses from the Planck+BK18 datasets. In~the $\Lambda$CDM+$r$+$\nu_0$ model, the~tensor-to-scalar ratio and the amplitude of the tensor spectrum are obviously influenced.

\noindent {\bf Acknowledgments}.
This work is supported by Natural Science Foundation of Shandong Province (grant No. ZR2021QA073) and Research Start-up Fund of QUST (grant No. 1203043003587).

%%%%%%%%%%%%%%%%%%%%%%%%%%%%%%%%%%%%%%%%
%%%%%%%%%%%%%%%%%%%%%%%%%%%%%%%%%%%%%%%%

%%%%%%%%%%%%%%%%%%%%%%%%%%%%%%%%%%%%%%%%
%%%%%%%%%%%%%%%%%%%%%%%%%%%%%%%%%%%%%%%%

%%%%%%%%%%%%%%%%%%%%%%%%%%%%%%%%%%%%%%%%
%%%%%%%%%%%%%%%%%%%%%%%%%%%%%%%%%%%%%%%%
\end{document}